\documentclass[12pt]{article}
\usepackage{authblk}

\usepackage[english]{babel}

\usepackage[a4paper,top=2cm,bottom=2cm,left=3cm,right=3cm,marginparwidth=1.75cm]{geometry}

\usepackage{amsmath,amsfonts,amssymb}
\usepackage{bm}
\usepackage{graphicx}
\usepackage[dvipsnames]{xcolor}
\usepackage[colorlinks=true, allcolors=blue]{hyperref}
\usepackage{pifont} 
\usepackage{siunitx} 
\usepackage[nameinlink]{cleveref}

\graphicspath{{epsfigures/}}

\title{Dynamical compartments in stirred tank reactors and Markov state modeling for mixing quantification: a transfer operator approach}

\author[1]{Anna Klünker}
\author[1]{Thanh Tung Thai}
\author[2]{Eike Steuwe}
\author[3]{Christian Weiland}
\author[3]{Yvonne Schade}
\author[2]{Alexandra von Kameke}
\author[1]{Kathrin Padberg-Gehle}

\affil[1]{Institute of Mathematics and its Didactics, Leuphana University Lüneburg}
\affil[2]{Heinrich-Blasius-Institute, Hamburg University of Applied Science}
\affil[3]{Institute of Multiphase Flows, Hamburg University of Technology}

\begin{document}
\maketitle
\begin{abstract}
  Identifying coherent flow structures in chemical reactors is crucial for understanding the mixing dynamics, which is essential for optimizing reactor performance. We demonstrate the use of a transfer operator method to find coherent flow structures such as almost-invariant sets and coherent sets, which are characterized by minimal mixing with the surrounding fluid, in a lab-scaled stirred tank reactor using both simulated and experimental Lagrangian trajectory data. The proposed method further enables a detailed analysis of the mixing behavior by computing expected residence times and mixing times. Additionally, a Markov-state-model describes the macroscopic transport dynamics between compartments in the reactor.
\end{abstract}

\section{Introduction}

Understanding and modeling the complex interactions between chemical reactions and fluid dynamics are key in the design and analysis of large-scale chemical reactors. Basic reactor models are global models that are based on reaction kinetics but assume the underlying fluid flow to be either perfectly mixed, ignoring spatial heterogeneities in substrate concentration, or completely segregated, observing each fluid element as a perfectly mixed stirred tank reactor without interaction with the neighboring fluid elements \cite{fogler_elements_2016}. In reality, a reactor is neither and the truth lies somewhere between these two assumptions \cite{levenspiel_chemical_1999} as a result of the interplay between efficiently mixed regions, dead zones, and bypasses. Being aware of these features is highly important for the design of resilient reactors and as such, in the development of SMART reactors in the Collaborative Research Centre 1615 ``SMART Reactors for Future Process Engineering'', funded by the German Research Foundation.

Computational Fluid Dynamics (CFD) is increasingly being used to simulate fluid dynamical processes at a high resolution to gain detailed insights into flow fields, including coherent flow patterns and mass transfer. Despite being very useful, CFD simulations are computationally demanding and have their limits when long-term studies of large-scale systems are carried out. The inclusion of reaction kinetics increases the complexity of the system even more and often leads to numerical instabilities.

Compartment models (CMs) build a bridge between basic reactor models and CFD. The reactor is considered as a network of ideally mixed zones with prescribed measured or precomputed exchange flows between these volumes, see \cite{Jourdan2019} for a recent review. In this way, CMs form a powerful framework that combines the advantages of local and global approaches: it is computationally feasible but nevertheless takes into account macroscopic heterogeneities in substrate concentrations resulting from limited global mixing. The major issue in the design of CMs is the determination of the compartments and the corresponding exchange flows between the compartments. Typically, CMs are derived from average or steady-state Eulerian flow field analysis, like average or mean velocity fields, average or mean turbulent quantities, or tracer concentrations \cite{Jourdan2019,cremer2020}. Automatic approaches using CFD snapshots have been proposed \cite{delafosse2014,Tajsoleiman2019,decarfort2024}, with recent methods exploiting unsupervised learning \cite{laborda2025,decarfort2026}.

As outlined above, a detailed insight into transport processes in the underlying flow field is key for the derivation of CMs. In fluid mechanics, these can be described from the Eulerian or Lagrangian point of view. Eulerian approaches consider instantaneous or time-averaged velocity fields to identify flow structures and compute exchange flows between different reactor volumes in terms of fluxes. The Lagrangian perspective, however, is directly linked to material transport as described by tracer trajectories and thus allows for a more precise, physically meaningful quantification of complex spatio-temporal transport patterns and pathways, which typically do not correspond to those identified by Eulerian methods, especially, when the Eulerian fluxes are merely derived from average fields as in the above cited literature.

To the best of our knowledge, none of the CM frameworks to date define compartments and quantify the exchange flows based on temporally resolved Lagrangian trajectory information, which may be obtained from  simulations but also from experimental time-resolved particle tracking in all three spatial dimensions (4D-PTV) \cite{HOFMANN2022, steuwe2026}. A very recent paper also tries to incorporate the Lagrangian aspect of the  task to derive the compartments \cite{decarfort2026}. In the cited work they compute the compartments in a stirred tank reactor (STR) directly from the concentration distributions of a passive tracer evolved in a frozen flow  field obtained by averaging a transient CFD. Here, our motivation is to develop an even simpler, but also inherently Lagrangian method to obtain these compartments directly from the objects that are being transported in the transient flow field, namely the Lagrangian trajectories of fluid parcels. The large advantage of our proposed approach in contrast to the former work \cite{decarfort2026} is that for the mixing time experiments no additional CFD simulation for the tracer concentration evolution has to be performed but rather, every possible initial position of the tracer feed  can be used for mixing time experiments with negligible additional effort as will be shown in the course of this work.

Lagrangian analysis was initially developed and is still in use for the study of transport patterns in geophysical flows \cite{Haller2015,prants2025}.  The concepts have been used to, e.g., explain plankton blooms before Madagascar \cite{Huhn2012} and oil spill dynamics \cite{Olascoaga2012,trinadha2024extraction}. Only recently these concepts were extended to chemical engineering \cite{Kameke2019,Llamas2020}. Numerous mathematical and computational methods originating from dynamical systems theory have been proposed to study Lagrangian transport and mixing processes \cite{Allshouse2015,Haller2023,Hadjighasem2017,Balasuriya2018,Badza2023}, building on different, but related notions and definitions of coherent flow structures. We will focus on the concept of almost-invariant and finite-time coherent sets \cite{dellnitz_junge_1999,FroylandLloydSan2010,Froyland2013,Karrasch2020}. These are regularly shaped fluid volumes that only weakly mix with their surroundings.
Spatially fixed regions in phase space with low escape probability, known as almost-invariant sets, were introduced within a probabilistic approach based on transfer operators \cite{dellnitz_junge_1999}. This ergodic theoretical concept was extended to the time-variant setting to identify coherent sets, which are mobile \cite{FroylandLloydSan2010,Froyland2013}, see also \cite{Froyland2015} for a related dynamic Laplacian framework.
The boundaries of such coherent regions can be identified within the geometric approach of Lagrangian coherent structures, where minimal curves or surfaces surround coherent sets \cite{Haller2015}.

When approximating the transfer operator within a set-oriented numerical framework, a stochastic transition matrix of a finite-state Markov chain is obtained, which provides the basis for Markov state models and empirical mixing studies. For the identification of the sets of interest, leading eigenvectors of the transition matrix are utilized. Extended methods address the study of almost-invariant and coherent sets that evolve, emerge or disappear \cite{Froyland2023,badza2024} and of bifurcations \cite{Blachut2020,Ndour2021}, whereas in \cite{kluenker2022open} the transfer operator approach was used to model and quantify mixing of two fluids. Recent data-based methods make use of unsupervised learning such as spatio-temporal clustering algorithms to extract coherent sets directly from Lagrangian trajectory data \cite{hadjighasem2016spectral,banisch2017understanding,Schlueter2017,padberg2017network, schneide_evolutionary_2022}.

In \cite{weiland_computational_2023} finite-time coherent sets were studied for the first time in the context of chemical process engineering, based on a Lattice-Boltzmann simulation of a \qty{2.8}{L} lab-scale STR. The dominant identified coherent flow structures were derived from Lagrangian tracer trajectories by spectral clustering and were found to be internally mixing. In \cite{schade2023msc} such coherent sets were tracked and observed in longer time intervals. A basic compartment model was derived that took into account the approximate location of the identified compartments and obtained the exchange rates between the compartment volumes from CFD. Finally, in \cite{steuwe2026} the coherent set computation from \cite{weiland_computational_2023} was repeated on experimentally measured data from 4D-PTV in a corresponding \qty{3}{L} STR, leading to results that are very similar to the simulated data.

In this contribution, we will push forward the Lagrangian studies of simulated and experimentally measured tracer trajectories in STRs with the perspective of designing a new generation of data-driven compartment models. We will apply for the first time the classical transfer operator-based approach for the identification of almost-invariant and finite-time coherent sets in a process engineering setting. Here, based on a grid discretization of the domain, we obtain the numerical transfer operator as a transition matrix of a finite-state Markov chain, where the transition probabilities are estimated by means of the tracer trajectories. The sets of interest, which form dynamical compartments, are identified from leading eigenvectors of that matrix. This allows for a Lagrangian determination of exchange rates and volume fluxes between those specified dynamical compartments. Additionally, mixing statistics and mixing times can be derived from the stochastic matrix. Finally, mixing time evaluations can be performed repeatedly choosing various possible initial concentration configurations with very moderate computational costs.

The remainder of the paper is organized as follows. In section \ref{sec:theory} we introduce the mathematical and computational background for our Lagrangian framework. After defining different notions of coherent sets that can serve as dynamical compartments (section \ref{sec:coherentsets}), we review the transfer operator-based approach and demonstrate how the corresponding Markov chain description of the underlying dynamics can be utilized for deriving transport and mixing quantities (section \ref{sec:set_oriented}).
The data sets for our computational studies, comprising both simulated and experimentally measured tracer trajectories from a lab-scale stirred tank reactor with two Rushton turbines and three baffles, are described in section \ref{sec:datasets}. In section \ref{sec:numericalresults} we present the results of our computational studies regarding the resulting compartments and Markov state models as well as mixing statistics. We conclude with a discussion and an outlook on current and future work in section \ref{sec:conclusion}.

\section{Theoretical background and methods}\label{sec:theory}
In the following, we describe the mathematical and computational background for coherent flow structures in a Lagrangian setting. We will sketch the general framework in the context of non-autonomous dynamical systems. To this end, let $M\subset \mathbb{R}^3$ be the physical space (in our setting the reactor domain), which we assume to be a compact subset of $\mathbb{R}^3$. We consider a time-dependent ordinary differential equation
\begin{equation}\label{eq:dgl}
  \dot{\bm{x}}=\bm{u}(\bm{x},t)
\end{equation}
with state (or position) $\bm{x} \in  M$, time $t \in \mathbb{R}$, and right-hand side $\bm{u}$, which, in our setting, corresponds to a velocity field in physical space and describes the advective motion of passive tracers in an incompressible fluid flow, i.e.\ volume is preserved.  $\bm{u}$ is assumed to be sufficiently smooth so that the flow map $\Phi: M \times \mathbb{R} \times  \mathbb{R} \to M$ exists. Given a tracer position $\bm{x}$ at time $t$, evaluating the flow map $\Phi(\bm{x}, t; \tau)$ gives its new position after a flow time $\tau$.

Note that in practice, the flow map $\Phi$ itself is not explicitly available, so we have to work with $N$ different Lagrangian tracer trajectories $(\bm{x}_i(t))_{t \in \mathbb{T}} \in \mathbb{R}^{d}$,  $i= 1,\ldots,N$. These may be obtained from the numerical simulation of \eqref{eq:dgl} or from experimental time-resolved particle tracking (4D-PTV). Such trajectories are only given for discrete measurement or evaluation times $t\in \mathbb{T} =\{t_0, t_1, \ldots, t_T\}$. We consider uniform time-steps $h$, i.e.\ $t_{k+1}=t_k+h$, $k=0, \ldots, T-1$, because this reflects the order of the datasets that are used in this study. Furthermore, for the derivation of the methods, we also assume for now that we do not have any gaps in observation. Taking into account the flow map $\Phi$, we obtain such time-discrete tracer trajectories $(\bm{x}_i(t))_{t \in \mathbb{T}}$, $i=1, \ldots,N$ from
\begin{equation}
  \bm{x}_i(t_{k+1}) = \Phi(\bm{x}_i(t_k), t_k; h).
\end{equation}
So, the given trajectories $(\bm{x}_i(t_k))_{t_k \in \mathbb{T}}$, $i=1, \ldots, N$, can be interpreted as the output of iterating the flow map for the $N$ initial conditions $\bm{x}_i(t_0)$.
\subsection{Almost-invariant and finite-time coherent sets}\label{sec:coherentsets}

We are interested in the identification of coherent flow structures in the Lagrangian frame of reference. Here, these structures are regular fluid volumes that are characterized by minimal mixing with the surrounding fluid. In the context of non-autonomous dynamical systems, such flow structures correspond to invariant, almost-invariant or finite-time coherent sets.

A set $A\subset M$ is called \textbf{$\Phi$-invariant} over $[t, t+\tau]$ if $\Phi(A, t; s)=A$ for all $0 \leq s\leq \tau$. That is, the set (or fluid volume) $A$ remains unchanged under the evolution of $\Phi$, i.e.\ it does not move as a whole but it might be internally mixing.

Almost-invariant sets obey an approximate invariance principle $\Phi(A, t; s)\approx A$ for all $0 \leq s\leq \tau$, see \cite{dellnitz_junge_1999} for mathematical details.
To be more precise, given the Lebesgue measure (volume measure) $\ell$ on $M$, we call a set $A\subset M$ with $\ell(A)\neq 0$ \textbf{almost-invariant} over the time span
$[t, t+\tau]$ if
\begin{equation}
  \label{eq:rhomu}    \rho(A):=\frac{\ell(A\cap\Phi(A,t;\tau))}{\ell(A)}\approx 1.
\end{equation}
If $A\subset M$ is almost-invariant over the interval $[t,t+\tau]$, then the proportion of $A$ that gets mapped to $A$ under the action of the flow map is close to 100\,\%. This means that a trajectory that started in $A$ remains in $A$ over the time-span $[t, t+\tau]$ with high probability.

Unlike almost-invariant sets that are fixed in phase space, \textbf{finite-time coherent sets} \cite{Froyland2013} are allowed to move under the evolution of the time-dependent system. The goal here is to find pairs of sets (or fluid volumes) $(A_t, A_{t+\tau})$, so that tracers starting in $A_t$ at time $t$ will be mapped to $A_{t+\tau}$ with high probability, i.e.\
  $A_{t+\tau} \approx \Phi(A_{t}, t; \tau)$. This translates into
  \begin{equation}\label{eq:rho}
    \rho(A_t, A_{t+\tau})=\frac{\ell(A_t \cap \Phi(A_{t+\tau}, t+\tau; -\tau))}{\ell(A_t)}  
  \end{equation}
  where $\Phi(A_{t+\tau}, t+\tau; -\tau)$ is a set that is obtained from mapping the set $A_{t+\tau}$ at time $t+\tau$ back to time $t$. Hence, equation \eqref{eq:rho} measures the proportion of the set $A_t$ at time $t$ that is mapped to the set $A_{t+\tau}$ at time $t+\tau$.
You can interpret a coherent pair as a fluid volume  that is observed as $A_t$ at time $t$ and as $A_{t+\tau}$ at time $t+\tau$, so that a large proportion of the tracers found in $A_t$ have moved to $A_{t+\tau}$ under the action of the flow and only very little tracers have been lost to other fluid volumes. A related interpretation is that if $A_t$ is an initial blob of high concentration of a chemical substance, then $A_{t+\tau}$ exhibits still most of this initial concentration.

Obviously this concept is ill-posed, because any pair $(A_t, A_{t+\tau})$ with $A_{t+\tau}:=\Phi(A_{t}, t; \tau)$ will be perfectly coherent in the sense that by definition all tracers in $A_t$ will be mapped to $A_{t+\tau}$, obtaining a maximal invariance ratio of $1$ in \eqref{eq:rho}.
To obtain well-posed problems for finding optimal sets (i.e.\ fluid volumes) that maximize the ratios \eqref{eq:rhomu} or \eqref{eq:rho} corresponding to almost-invariance and finite-time coherence, respectively, some regularity and mass constraints have to be imposed. These require that the sets of interest are ``regular'' in the sense that they are stable under small perturbations such as diffusion \cite{dellnitz_junge_1999,Froyland2013}. This will be ensured by our computational framework approach, which includes numerical as well as explicit diffusion (see section \ref{sec:set_oriented}). This robustness with respect to perturbation means in our case, that scalar quantities such as concentrations would stay contained in the coherent fluid volumes in the sense that there will only neglegible advective-diffusive transport to the surrounding fluid. Thus, the almost-invariant and finite-time coherent sets considered here are fluid volumes that are transported and maybe slightly deformed and possibly well-mixed internally but that resist mixing with the other fluid initially outside of this volume.

We will not focus on the analytical constructions \cite{Froyland2013,froyland_padberg_2014almost} in this work, but rather on data-driven, computational and applied aspects of coherent flow structures, especially in process engineering problems. In the following, we revisit the set-oriented numerical framework for approximating almost-invariant and finite-time coherent sets. We adapt this framework in view of the given simulated or measured trajectories.
\subsection{Set-oriented method based on transfer operators} \label{sec:set_oriented}

As we are interested in the transport of sets and densities under the action of the flow map, we now review a corresponding mathematical concept that is tailored to this problem. We consider the transfer or Perron-Frobenius operator $\mathcal{P}_{t,\tau}: L^1(M,\ell) \to L^1(M,\ell)$  associated with the flow map $\Phi$. As our system is assumed to be incompressible, this linear operator is simply defined by
\begin{equation}\label{eq:P}   \mathcal{P}_{t,\tau}f(\bm{x})=f(\Phi(\bm{x}, t+\tau; -\tau)).
\end{equation}
The interpretation is that if $f$ is a density (e.g.\ representing a scalar field like a concentration field on $M$) and $f(\bm{x})$ the density value at tracer position $\bm{x}$ at time $t$, then $\mathcal{P}_{t,\tau}f(\bm{x})$ describes the density value in $\Phi(\bm{x}, t; \tau)$ at time $t+\tau$ induced by the flow map.
In \cite{Froyland2013,froyland_padberg_2014almost} it was shown that maximizing  $\rho$ in \eqref{eq:rhomu} and \eqref{eq:rho} can be described in the framework of optimizing an inner product involving a compact self-adjoint operator obtained from $\mathcal{P}_{t,\tau}$. We refer to these works for the functional-analytical background.

We assume for all that follows that our domain $M$ is compact and invariant, i.e.\ no trajectories will leave $M$.  This is a matching assumption given that the trajectory data studied here originates from simulations and measurements of a closed system. However, the method can be extended to deal with non-invariant subsets of $M$\,\cite{froyland_santi_monahan_10,Froyland2013,froyland_padberg_2014almost}, which is useful for the future consideration of continuous reactor operations of the STR (CSTR).
\subsubsection{Numerical approximation of transfer operators}\label{sec:numapproxTO}

Let $\{B_1, \ldots, B_n\}$ be a partition of $M$, this is a set of $n$ many small subdomains of the reactor domain so that $\bigcup_{i=1}^n B_i =M$. Here, the partition elements are Cartesian grid cells (``boxes''), but other settings are possible. Applying Ulam's method \cite{ulam} a finite-rank approximation of $\mathcal{P}_{t,\tau}: L^1(M, \ell) \to L^1(M,\ell)$ is given via the transition matrix
\begin{equation}\label{eq:pij}
  P_{ij}=\frac{\ell(B_i\cap \Phi(B_j, t+\tau;-\tau))}{\ell(B_i)}, \; i,j=1,\ldots, n
\end{equation}
where we drop the $t$ and $\tau$-dependence of $\bm{P}$ for brevity. In practice, we count how many of the particles released in $B_i$ end in $B_j$. For this,
we consider uniformly distributed sample points $\bm{z}_{i,r}, r=1,\ldots,R$ chosen in each partition element $B_i$, $i=1,\ldots,n$. The entries $P_{ij}$ of the transition matrix $\bm{P}$ are then estimated via
\begin{equation}\label{eq:pij_approx}
  P_{ij}\approx\frac{\#\{r:\Phi(\bm{z}_{i,r}, t; \tau)\in B_j\}}{R},
\end{equation}
where $\#$ denotes cardinality, i.e.\ the number of elements in a set. $\bm{P}$ is a sparse, row-stochastic matrix and thus all its eigenvalues are contained in the unit circle. For efficient computation of the transition matrix $\bm{P}$ we use the GAIO library \cite{GAIO}.

The dynamics induced by $\bm{P}$ can be interpreted as follows: if $\bm{p}\ge 0$ (component-wise) is a density vector (e.g.\ containing the values of a scalar quantity, such as a concentration, for each box), then $\bm{p}'=\bm{p}\bm{P}$ is the push-forward of $\bm{p}$ under the discretized action of the flow map $\Phi(\cdot, t;\tau)$ and the vector $\bm{p}'$ describes the evolved values after a time span $\tau$. Since $\bm{P}$ is row-stochastic, the total sum of the entries of $\bm{p}$ is preserved under this evolution.
In particular, if $\sum_i p_i = 1$, then $\bm{p}$ is a probability vector, and $\bm{p}'$ remains a probability vector.
Note that the numerical scheme introduces 
diffusion in the order of magnitude of the box sizes, which is also theoretically needed for robust results \cite{Froyland2013,froyland_padberg_2014almost}.

\begin{figure}[!htb]
\centering
\includegraphics[width=0.45\textwidth]{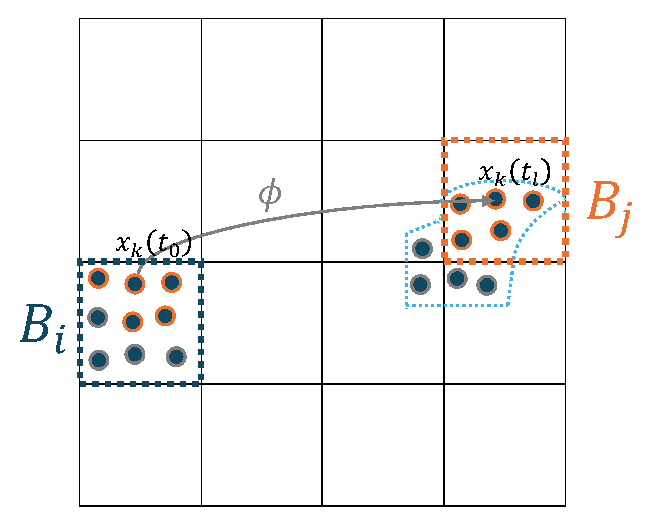}
\caption{Schematic representation of the transition matrix construction according to equation \eqref{eq:pij_approx_traj}: Each entry $P_{ij}$ is estimated as the fraction of particles from $B_i$ that end in $B_j$ (circled in orange).}
\label{fig:schema_transitionmatrix}
\end{figure}

To construct the numerical transfer operator from given time-discrete tracer trajectories $(\bm{x}_i(t))_{t \in \mathbb{T}}$, $i=1, \ldots, N$, $\mathbb{T} =\{t_0, t_1, \ldots, t_T\}$, where $t_{j+1}=t_j+h$, one has to adapt \eqref{eq:pij_approx} as follows. First one has to fix a flow time $\tau = l\cdot h$ as a multiple of $h$, which is the time step of the trajectory data, and considers the respective interval $[t_0, t_{l}]$ with $t_0, t_{l}=t_0+\tau \in \mathbb{T}$.
Let $I_i\subset \{1,\ldots, N\}$ be the $R_i$ indices of those initial conditions that can be found in box $B_i$, i.e.\ $\bm{x}_k(t_0)\in B_i$, $k \in I_i$. $P_{ij}$ then gives the proportion of these initial conditions that will be mapped to box $B_j$ (see Figure \ref{fig:schema_transitionmatrix} for an illustration):
\begin{equation}\label{eq:pij_approx_traj}
  P_{ij}\approx\frac{\#\{ k \in I_i: \bm{x}_k(t_l)\in B_j\}}{R_i}.
\end{equation}
The given data may pose serious constraints. Often the trajectories are nonuniformly distributed and the number of data points per box at the initial and/or the final time step is not sufficient to represent the transition probabilities well. Although the computational framework already incorporates numerical diffusion in the order of magnitude of the box size, we can smooth our construction by introducing additional diffusion. A simple way is to consider a multivalued mapping approach: We introduce $N_{\text{diff}}$ diffusive duplicates of each tracer $\bm{x}_k(t_l)$ at the final time step $t_l$. For this we generate a grid in an $\epsilon$-ball (a ball with radius $\epsilon$) around each $\bm{x}_k(t_l)$. The $N_{\text{diff}}$ grid points are then positions of pseudo tracers at time $t_l$ with the same initial position $\bm{x}_k(t_0))$ as the original tracer. Let $R_i$ be the number of tracers starting in box $B_i$ and $\hat{R}_{ij}^{\epsilon}$ the number of diffused pseudo tracers starting in box $B_i$ and ending in box $B_j$. The pseudo tracers around each real tracer particle start in the same location as the real particle but end up at slightly moved locations, mimicking a diffusive process. Then we get the following diffused transition matrix
\begin{equation}\label{eq:pij_approx_traj_diff}
P_{ij}^{\text{diff}}\approx\frac{\hat{R}_{ij}^{\epsilon}}{N_{\text{diff}}\cdot R_i}.
\end{equation}
To study the transport over a series of adjacent time intervals $T_1=[t_0, t_l]$, $T_2=[t_l, t_{2l}], \ldots ,T_f=[t_{(f-1)l}, t_{fl}]$ of length $\tau=l\cdot h$, where $t_{k\cdot l}\in \mathbb{T}$ for $k=0, \ldots, f$, we can construct a family of transition matrices $\bm{P}_{t_0,\tau}$, $\bm{P}_{t_l,\tau}, \ldots$ over the respective time intervals. For a time-dependent velocity field $\bm{u}$, these matrices are different in general. When the system is autonomous, i.e., $\bm{u}$ is time independent, $\bm{P}_{t_{k\cdot l},\tau}$ only depends on the length $\tau$ of the intervals and not on the initial times. Therefore, in theory, a series of time intervals with equal length would lead to the same transition matrices (time-homogeneous Markov chain). Further, when $\bm{u}$ is a time-dependent periodic velocity field with period $\tau$, that
is, $\bm{u}(\bm{x}, t) = \bm{u}(\bm{x}, t + \tau)$, and the length of all time intervals is $\tau$, we would obtain also the same transition matrix for all such adjacent  intervals.
For autonomous and periodic systems this can be exploited to enrich the database: We can crop the given trajectories into segments of length $\tau$ and use all of these to construct a transition matrix on $T_1=[t_0, t_l]$.
\subsubsection{Extraction of almost-invariant and finite-time coherent sets}\label{sec:extractAIS}
The general aim is to find an optimal partition of $M$ into subsets (i.e.\ compartments) $A_1, \ldots, A_{K+1}$, that mix little with the rest. This processing step can also be understood as a means to finding the best number of compartments and their shape in order to represent the dynamic system of transport in a CM model. The number and shape of the ideal compartmentalization is current subject of study \cite{decarfort2026}. In our computational framework each $A_j$, $j=1, \ldots, K+1$, is made up of boxes, i.e.\   $A_j = \bigcup_{i \in I_{A_j}}B_i$, where $I_{A_j}$ denotes the set of indices $I_{A_j}\subset \{1, \ldots, n\}$ of the grid cells that form $A_j$.  We obtain such a partition based on the numerical transfer operator $\bm{P}$.  In a nutshell, the transition matrix $\bm{P}$ is a stochastic matrix, with eigenvalues contained in the unit circle. In practice, we will form a transition matrix $\bm{R}$ of a reversible Markov chain from $\bm{P}$ \cite{froyland_padberg_2014almost}, which exhibits a real-valued spectrum. The leading eigenvalue $\lambda_1$ is always 1 and the corresponding eigenvector $\bm{v}_1$ is constant, while all further eigenvectors to eigenvalues  are orthogonal to $\bm{v}_1$ and thus contain signed entries. One would find a nearly optimal partition of $M$ into two almost-invariant sets from the sign structure of the second eigenvector $\bm{v}_2$ to the eigenvalue $\lambda_2<1$ (but close to $1$) by defining $A_1=\bigcup_{\{j: v_{2,j}\geq 0\}} B_j$ and $A_2=\bigcup_{\{j: v_{2,j}< 0\}} B_j$ \cite{dellnitz_junge_1999}. In general, almost-invariant sets are extracted from several eigenvectors $\bm{v}_2, \ldots, \bm{v}_K$ of this transition matrix $\bm{R}$ obtained from $\bm{P}$ to eigenvalues $1>\lambda_2\geq \ldots \geq \lambda_K$, where $K$ is chosen such that $\lambda_K- \lambda_{K+1}$ is large (i.e.\ there is a spectral gap). Similarly, for finite-time coherent sets singular vectors of $\bm{P}$ or a related matrix are considered. We state the exact algorithms in appendix \ref{sec:appendix_algorithms}. A strong improvement in the capturing of the mass transfer dynamics by adding further compartments can only be expected when the eigengap between successive eigenvalues is high.

To extract almost-invariant and coherent sets from the corresponding vectors $\bm{v}_2, \ldots, \bm{v}_K$, we post-process these vectors to disentangle the sets using the SEBA (Sparse EigenBasis Approximation) algorithm \cite{froyland_sparse_2019}:
The basis $\bm{V} := (\bm{v}_1, \dots, \bm{v}_K)$ of the eigenspace $\mathcal{V} \in \mathbb{R}^{n}$ is transformed into a sparse basis $\bm{S} :=(\bm{s}_1, \dots, \bm{s}_K)$ of the vector subspace $\mathcal{S}$, such that $\mathcal{V} \approx \mathcal{S}$. The entry $s_{ij}$ of the matrix $\bm{S}$ gives the likelihood of box $B_i$, $i=1, \ldots, n$ to belong to the box-valued sets $A_j$, $j=1, \ldots, K$ as defined above. By choosing a threshold, each box $B_i$ can be assigned to a specific compartment $A_j$ or to the incoherent background, which will form the additional compartment $A_{K+1}$, respectively \cite{froyland_sparse_2019}. The vector $\bm{s}_{\text{max}} \in \mathbb{R}^{n}$, which describes an overall cluster likelihood for each particle with entries $s_i = \text{max}_j(s_{ij})$, $i=1, \ldots, N$, $j=1, \ldots, K$ can also be considered. 
\subsubsection{Markov chain statistics}\label{sec:mixingstatistics}
Using a transition matrix $\bm{P}$, we can evolve any (box-discretized) initial distribution in time. In particular, given a family of transition matrices over adjacent time intervals, $T_1$, $T_2$, \ldots, $T_f$ we can push forward an initial density vector $\bm{p}$ via $\bm{p}\bm{P}_{t_0,\tau}\bm{P}_{t_l,\tau}\cdot \ldots \cdot \bm{P}_{t_{(f-1)l},\tau}$. The transfer operator over a long time span can thus be approximated by  multiplication of transition matrices computed over shorter time intervals. In practice, only consecutive matrix-vector multiplications with sparse transition matrices have to be carried out, which can be computed very efficiently.  Hence, fluid mixing can be described and quantified (e.g.\ by the variance or other mixing measures) using a numerical transfer operator $\bm{P}$ \cite{kluenker2022open}.

In the following, we assume that we have a time-homogeneous Markov chain with transition matrix $\bm{P}$, corresponding to a time-periodic or autonomous velocity field.
We consider a set $A$ consisting of boxes $B_i$, $i\in I_A$, where $I_A \subset \{1, \ldots, n\}$ is the set of indices such that $\bigcup_{i\in I_A} B_i =A$. We can compute the \textbf{expected residence times} for $A$, the expected time to stay in $A$ before leaving and reaching any other part of the domain $M\setminus A$.
 Let $\bm{P}_A $ be the transition matrix $\bm{P}$ restricted to the states in $A$. We assume that $A$ is transient, i.e.\ eventually all mass initialized in $A$ will leave $A$ under the repeated action of $\bm{P}_A$. 
The expected residence time to stay in $A$ when starting in box $B_i\subset A$ is then given as
\begin{equation}\label{eq:residencetimes}
r_i=\sum_{j\in I_A} (\bm{I}-\bm{P}_A)^{-1}_{ij}.
\end{equation}
In practice, this can be computed by solving
\begin{equation}
(\bm{I}-\bm{P}_A) \bm{r}=\bm{1}, \label{eq:SLEresidencetimes}
\end{equation}
where the vector $\bm{r}$ contains the entries $r_i$, and $\bm{1}$ denotes the all-ones vector.

For a given initial density vector $\bm{v}$ (for example an initial blob of high concentration), the \textbf{mixing time} describes how long it takes for the system to approach equilibrium within a defined tolerance. We evolve $\bm{v}$ until 95~\% of all boxes have a concentration value within ±5~\% of the equilibrium value (given by the leading left eigenvector of $\bm{P}$), and define the number of time steps required as the mixing time. In the following, both the expected residence time and mixing time are given in number of discrete time steps of length $\tau$.

Further, we can set up a \textbf{Markov state compartment} model by considering the transitions between almost-invariant sets and the surrounding region.
Suppose you are given the $K+1$ disjoint sets of interest $A_1$, \ldots, $A_{K+1}$ that are composed of boxes and $\bigcup_{k=1}^{K+1} A_k =M$.
Let $I_{A_1}$, \ldots , $I_{A_{K+1}}$ denote the sets of indices such that
$\bigcup_{i \in I_{A_k}} B_i =A_k$, $k=1, \ldots, K+1$. Note that we think of $K$ almost-invariant sets plus the remainder of the domain, which will form the set $A_{K+1}$. 
The transition matrix $\bar{\bm{P}} \in \mathbb{R}^{K+1,K+1}$ that describes the coarse grained dynamics between these $K+1$ compartments is obtained as follows:
Let $\bm{e}_{A_k} \in \mathbb{R}^n$ be an indicator vector of the set $A_k$, i.e.\ $(\bm{e}_{A_k})_i = 1$ for $i\in I_{A_k}$ and $=0$ otherwise. The entries of $\bar{\bm{P}}$ are computed via
\begin{equation}
\bar{P}_{ij} =\frac{\bm{e}_{A_i}^\top \bm{P} \bm{e}_{A_j}}{\#I_{A_i}},  \; \; \; i,j =1, \ldots, K+1.\label{eq:coarseP}
\end{equation}
In case of finite-time coherent sets as compartments, one considers $K+1$ different pairs of sets $(A_{k, t},A_{k, t+\tau})$, $k=1, \ldots, K+1$, with indicator vectors $\bm{e}_{A_{k,t}}$ and $\bm{e}_{A_{k,t+\tau}}$, respectively, where again the $(K+1)$-th pair is identified with the incoherent background. The entries of the coarse-grained
transition matrix $\bar{\bm{P}}_{t,\tau} \in \mathbb{R}^{K+1, K+1}$ are then computed from $\bm{P}_{t,\tau}$
in analogy to equation \eqref{eq:coarseP} as
\begin{equation}
  \bar{P}_{ij,t,\tau } =\frac{\bm{e}_{A_{i, t}}^\top \bm{P}_{t,\tau} \bm{e}_{A_{j, t+\tau}}}{\#I_{A_{i,t}}},  \; \; \; i,j =1, \ldots, K+1.\label{eq:coarseP_time}
\end{equation}
This equation gives the conditional probability that a tracer initialized in compartment $A_{i,t}$ at time $t$ can be found in compartment $A_{j, t+\tau}$ at time $t+\tau$. The probability $\bar{P}_{ii,t,\tau}$ describes the transition from $A_{i,t}$ to $A_{i,t+\tau}$, i.e.\ the probability to stay in the (time-dependent) coherent compartment, which should be large by construction of coherent sets. 

These Markov state models provide simplified representations of the macroscopic dynamics of the system. Typically, there is little transport between different compartments, which is how these sets are characterized from equations \eqref{eq:rhomu} and \eqref{eq:rho}, respectively, while there can be a high level of mixing within individual compartments.

Using the volumes of the compartments, one can also compute the \textbf{transition volumes} or volume flow rates $Q$ from $\bar{\bm{P}}$ (similarly for $\bar{\bm{P}}_{t,\tau}$).
The entry $\bar{p}_{ij}$ is the conditional transition probability from compartment $A_i$ to $A_j$. The volume flow rate from compartment $A_i$ to $A_j$ is then given as
\begin{equation}
  q_{ij}=\frac{\bar{p}_{ij} \cdot \ell(A_i) }{\tau},
\end{equation}
where the volume $\ell(A_i)=\sum_{j\in I_{A_i}}\ell(B_j)$ can be straightforwardly computed from the side lengths of the boxes $B_i$.
If the geometric volumes of the boxes $B_j$ are only approximate (for example, if the boxes are cubic and extend partially outside the actual reactor), one can correct the compartment volumes using the leading left eigenvector $\boldsymbol{\pi}$ of ${\bm{P}}$, normalized to match the total volume. One estimates the volume of compartment $A_i$ then as $\tilde{\ell}(A_i) = \sum_{j\in I_{A_i}} \pi_j \ell(B_j)$.

\section{Stirred tank reactor data}\label{sec:datasets}
\begin{figure}
\centering
\includegraphics[width=5cm]{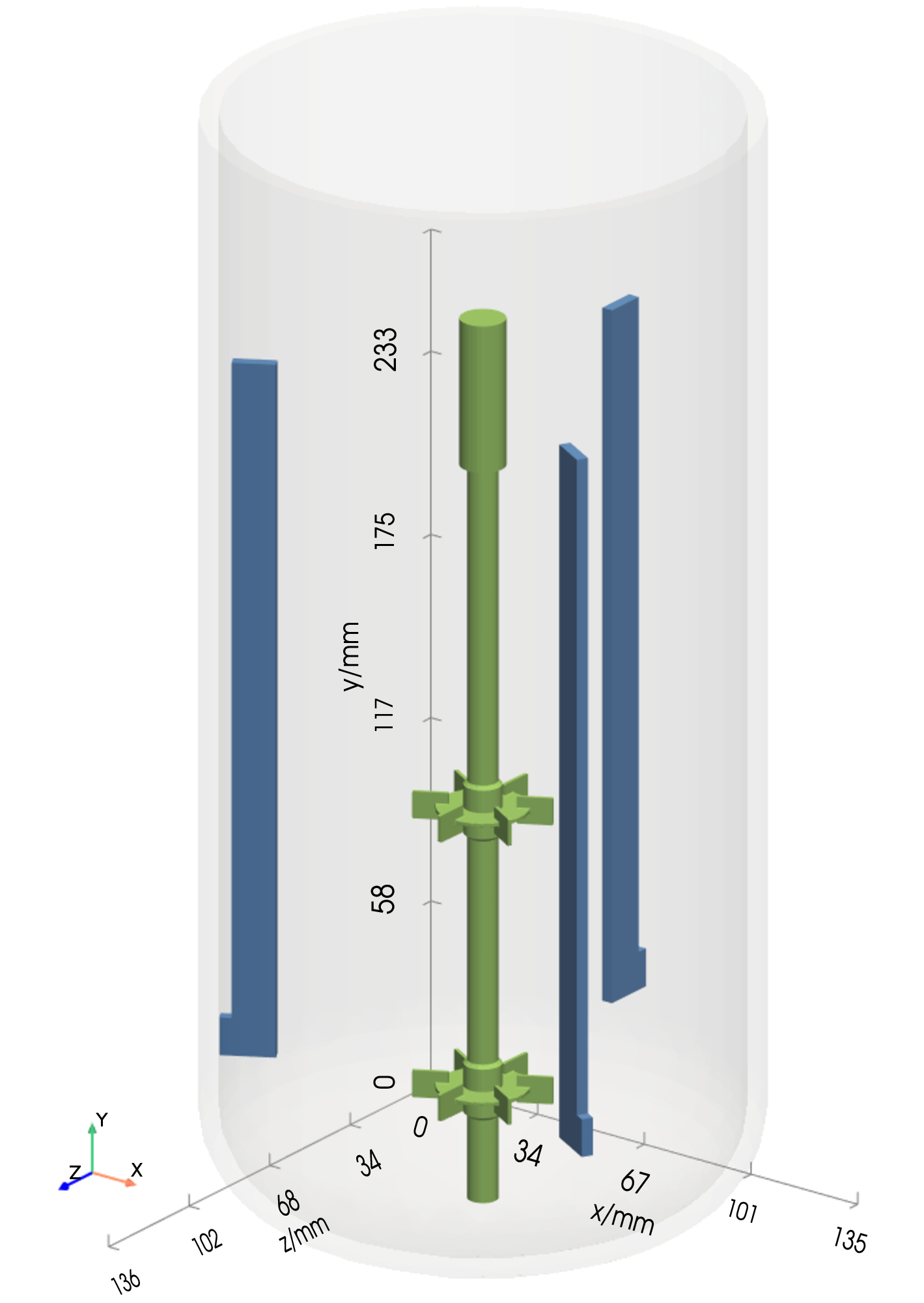}
\caption{Schematic representation of the stirred tank reactor. Dimensions are: length (x): $127.94 \,\text{mm}$, width (z): $127.94 \,\text{mm}$, height (y): $223.64 \,\text{mm}$.}
\label{fig:str_container}
\end{figure}

The trajectory data of the STR originate from both simulations and experiments conducted in a lab-scaled vessel (see Figure \ref{fig:str_container}) filled with water and stirred by two Rushton turbines at 252~rpm. The STR is further equipped with three baffles and a korbbogen head bottom.
\paragraph{Simulated data}
The data results from Lattice Boltzmann simulations conducted with the program M-Star CFD version 3.7.123. A total of $g = 300$ lattice sites are chosen across the vessel's diameter, resulting in a resolution of $\Delta x = \qty{4.3e-4}{m}$.
This resolution was found to result in a grid independent simulation in previous publications \cite{weiland_computational_2023}. To account for turbulent modeling, the simulations are carried out as Large Eddy Simulation (LES). The turbulent viscosity is modeled with the Smagorinsky-Lilly subgrid-scale model \cite{smagorinsky_general_1963} with a Smagorinsky constant of $C_\text{S}=0.1$. The simulation is run for a time of 60 seconds to achieve a quasi-stationary flow. After this time, the volumetric data concerning the time-dependent velocity field is stored at a frequency of $f_\text{save} = \qty{500}{\hertz}$ over a period of approximately $t_\text{save} = \qty{2.5}{\s}$, covering slightly more than ten full stirrer rotations. This data is then used to advect a total of roughly \num{342,000} particles by the flow to retrieve individual trajectories. A more detailed description of this process is given in \cite{weiland_computational_2023, schade2023msc}.
\paragraph{Experimental dataset}
The experimental dataset originates from a 4D Particle Tracking Velocimetry (4D-PTV) measurement that was carried out and published by Steuwe et al.\ \cite{steuwe2026}. 4D-PTV is a novel measurement technique in which particles are tracked in a flow using multiple cameras with different lines of sight, in order to obtain time-resolved reconstructions of the particle tracks in three-dimensional space. From these trajectories the Lagrangian velocity can be derived by differentiation \cite{schanz_shakethebox_2016}. In the study by Steuwe et al.\ \cite{steuwe2026}, the experiment was conducted at \qty{22}{\celsius} in a handmade glass vessel filled with \qty{3}{\liter} of bidistilled water and seeded with approximately 40,000 fluorescent, monodisperse particles measuring \qty{50}{\micro\meter} in diameter. Due to their small size compared to the underlying flow structures and their neutral buoyancy, the particles are considered to behave as passive tracers in the flow \cite{steuwe2026}. The experimental setup includes three baffles and two Rushton turbines, operating at \qty{252}{rpm}. This matches the setup for the simulation dataset by Weiland et al. \cite{weiland_computational_2023} introduced in the previous section. The resulting dataset consists of $37,879 \pm 13$ particles per timestep tracked over 2,818~timesteps at a rate of \qty{500}{\Hz}. This corresponds to a total investigation time of \qty{5.64}{\second}. Since the experiment was carried out using five high-speed cameras positioned around the entire STR, the dataset covers the full reactor volume. Due to the 4D reconstruction algorithm, the tracks are occasionally lost and found again (less than 100 particles are lost and found each time step), resulting in a total of 216,000 individual particle tracks. The loss of particle trajectories must be taken into account when using the analysis method presented here.
\section{Computational results}\label{sec:numericalresults}
In this section we carry out computational 
analyses in a STR using the simulated and experimental trajectory data as introduced in section \ref{sec:datasets}, applying the set-oriented framework discussed in section \ref{sec:set_oriented}.
We demonstrate the extraction of compartments (i.e.\ almost-invariant and coherent sets) using the transfer operator approach within a set-oriented numerical framework. Moreover, we derive Markov state models to describe the macroscopic dynamics and study scalar mixing.
\subsection{Compartments and Markov state model}
First, we apply our set-oriented method to ten consecutive stirrer rotations of the simulated data, i.e.\ $\tau\approx\qty{0.2381}{\s}$ (one rotation).
We consider the respective parts of the simulated trajectories on the ten different time intervals $[0,1], [1,2], \ldots [9,10]$ (time expressed in number of rotations). To set up the transitions matrices, the domain is partitioned into small boxes with side lengths $(l_{Bx},l_{By},l_{Bz})=(4.0,3.8,4.0)$\,mm.
To compute the transition matrices, we add diffusion (radius $\epsilon=\qty{4}{\mm}$, a regular grid containing 33 test points) as described in equation \eqref{eq:pij_approx_traj_diff} in section \ref{sec:set_oriented}. The number of boxes ranges between 47,993 and 49,029 for the different time intervals, since we can only use boxes with available trajectory data.

\begin{figure*}[htb] \centering
\includegraphics[width=1\textwidth]{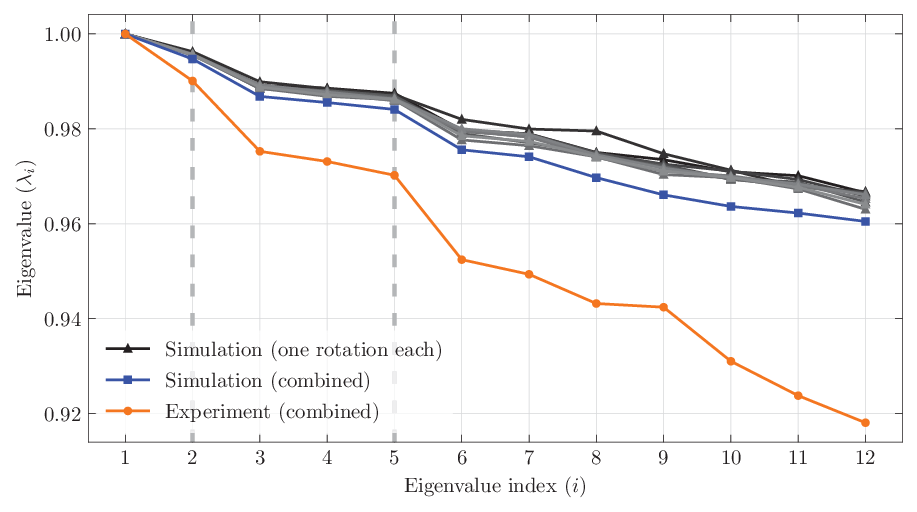}
\caption{Leading 12 eigenvalues for ten symmetrized transition matrices for the time intervals $[0,1], [1,2], \ldots [9,10]$ (time expressed in number of rotations) for the simulated data (\textbf{\textcolor{black}{\ding{115}}}, \ldots ,\textbf{\textcolor{gray}{\ding{115}}}). Combined transition information for one stirrer rotation into a single transition matrix: Eigenvalues of the symmetrized transition matrix for the simulated (\textbf{\textcolor{blue}{\ding{110}}}) and the experimental (\textbf{\textcolor{orange}{\ding{108}}}) data. } \label{fig:simexpdata}
\end{figure*}

We use Algorithm \ref{appendix:AIS} in the appendix and compute the leading eigenvalues of the symmetrized transition matrices (Figure \ref{fig:simexpdata}, grey curves). For all matrices, there is a clear eigengap after the fifth eigenvalue indicating the existence of $K=5$ prominent almost-invariant sets. For some of the matrices there are further spectral gaps, e.g.\ after the eighth or eleventh eigenvalues, but not as clearly.
Therefore, further partitioning of the reactor into more compartments is not necessary to grasp the large-scale chaotic transport for the studied STR with two Rushton stirrers. In comparison to other work on STRs \cite{decarfort2026,Maldonado2025}, the determined number of necessary compartments is very low, however, here the shape of the compartments was determined from the Lagrangian information and not from a frozen flow field that does not represent the dynamical system  for transient flows. 

\begin{figure*}[ht!]
\includegraphics[width=1\textwidth]{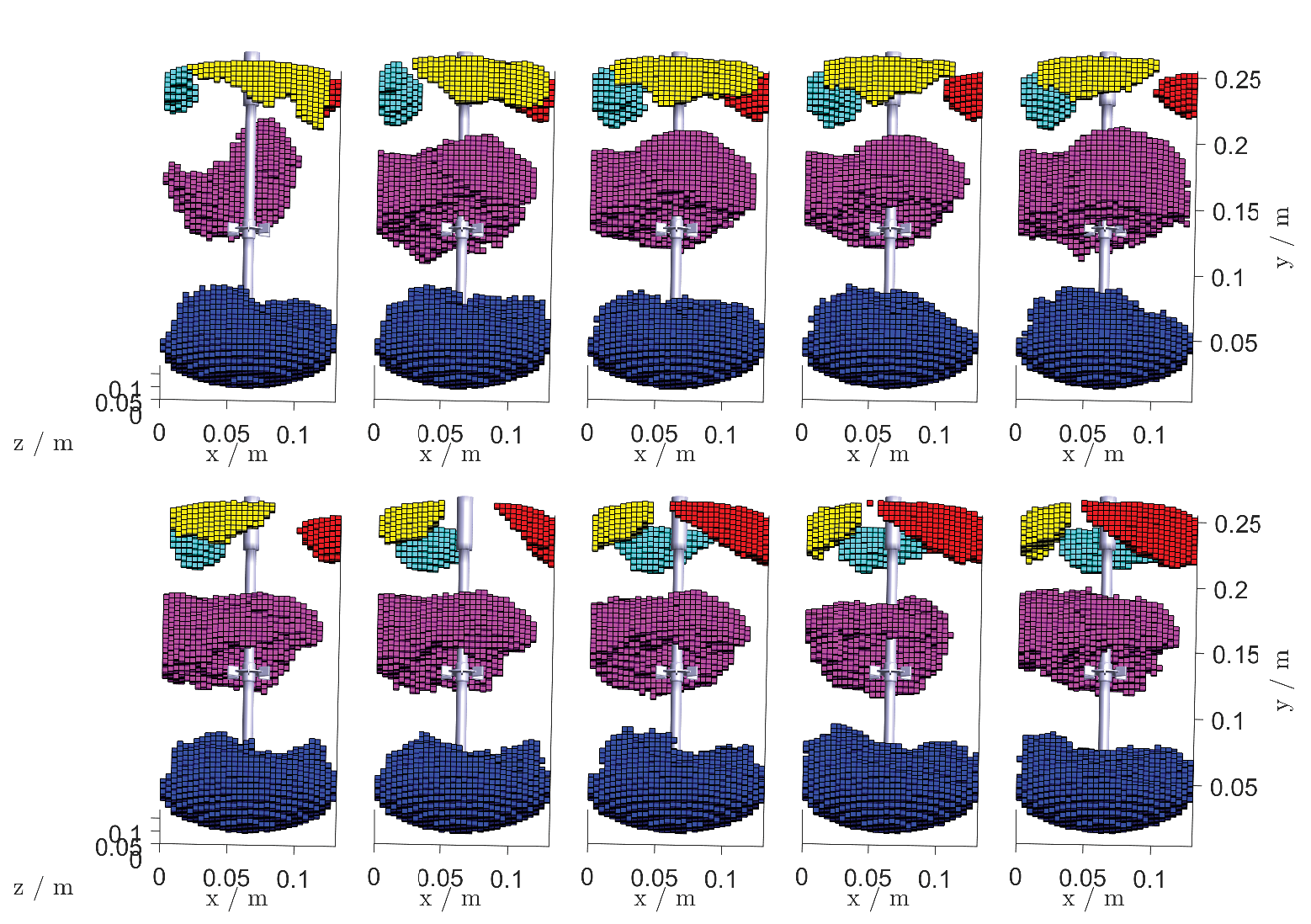}
\caption{Extracted five almost-invariant sets on time intervals $[0,1], [1,2], \ldots [9,10]$ with time expressed in number of rotations.}\label{fig:tenalmost}
\end{figure*}

To extract the almost-invariant sets, we consistently use a cut value of 0.7 on the SEBA vectors (post-processed eigenvectors, see section
\ref{sec:extractAIS}).
Figure \ref{fig:tenalmost} shows the extracted five almost-invariant sets for ten different stirrer rotations. In each of the ten cases we obtain three smaller regions located at the top, one in the center, and one at the bottom of the reactor, which appear to be very similar in shape and position for the ten different stirrer rotations. In particular, the bottom set seems to be very robust, whereas one observes a slight rotation of the upper three sets from one time interval to the next.

\begin{figure*}\centering
\begin{tabular}{@{}ccc@{}}  \includegraphics[width=0.3\textwidth]{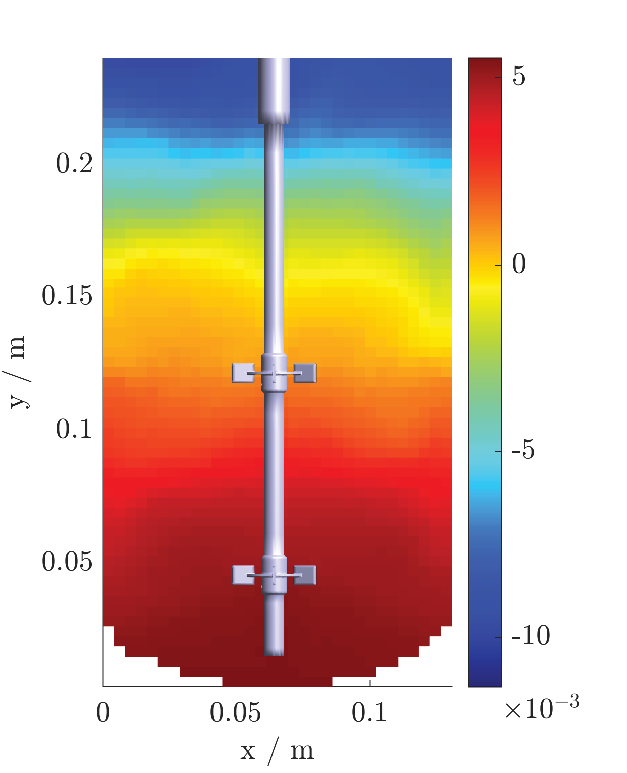} &  \includegraphics[width=0.3\textwidth]{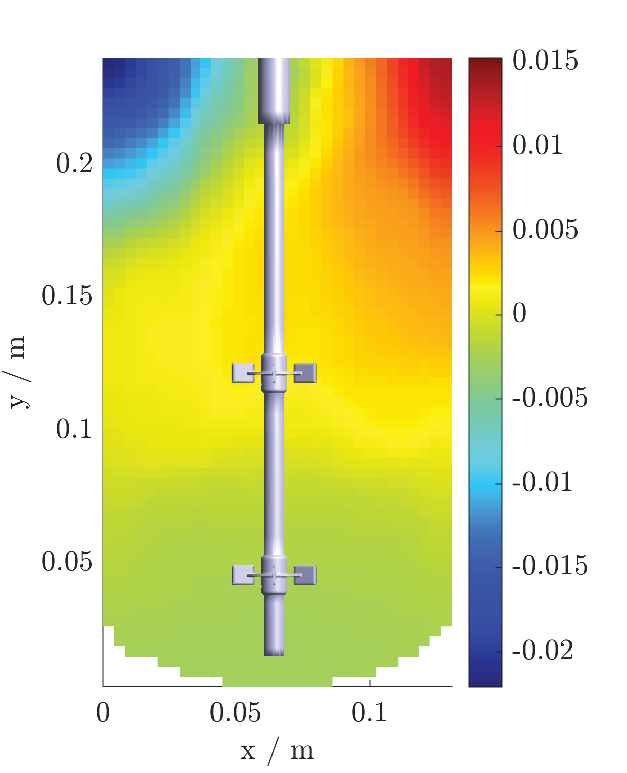} &  \includegraphics[width=0.3\textwidth]{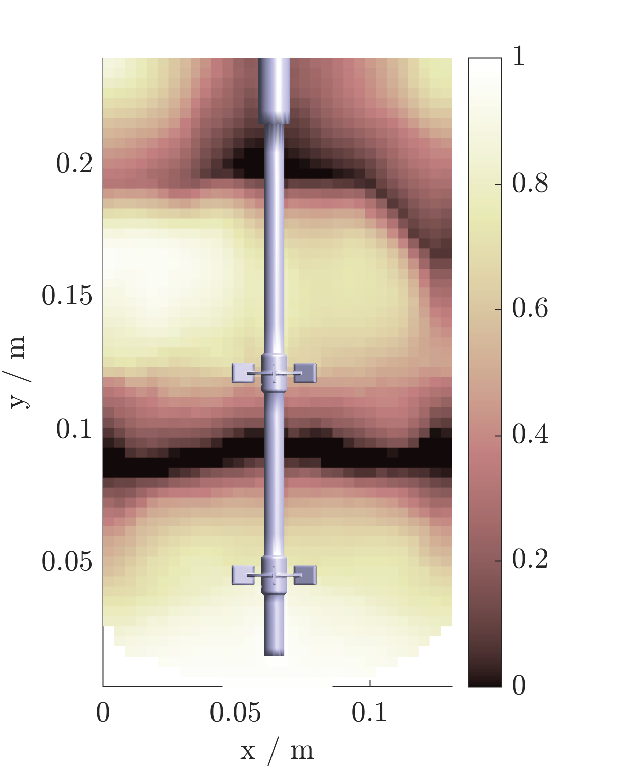} \\
  {\scriptsize(a)} & {\scriptsize (b)} & {\scriptsize (c)} \\
\end{tabular}
\caption{Eigenvectors to the second (a) and fourth (b) eigenvalue of the symmetrized transition matrix (see appendix \ref{appendix:AIS}) computed from the combined transition information for one stirrer rotation and $\bm{s}_{max}$ (c) serving as a cluster indicator; all shown in the middle plane $z=\qty{0.0635}{\m}$.} \label{fig:eigenvec}
\end{figure*}

\begin{figure*}[!htb]\centering
\begin{tabular}{@{}cc@{}} \includegraphics[width=0.4\textwidth]{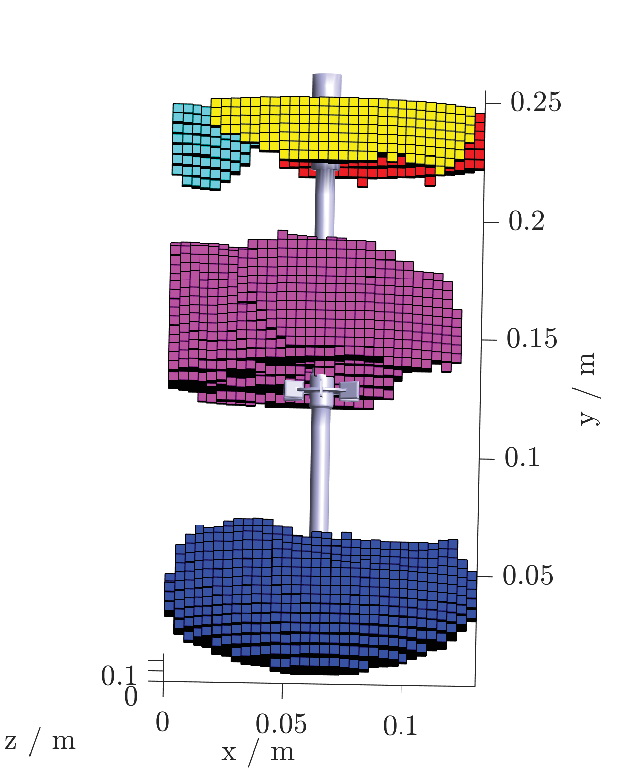} &  \includegraphics[width=0.55\textwidth]{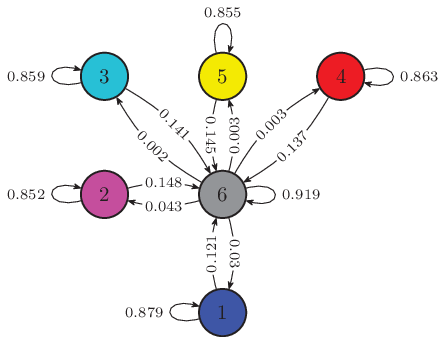} \\
  {\scriptsize(a)} & {\scriptsize (b)} \\  \includegraphics[width=0.4\textwidth]{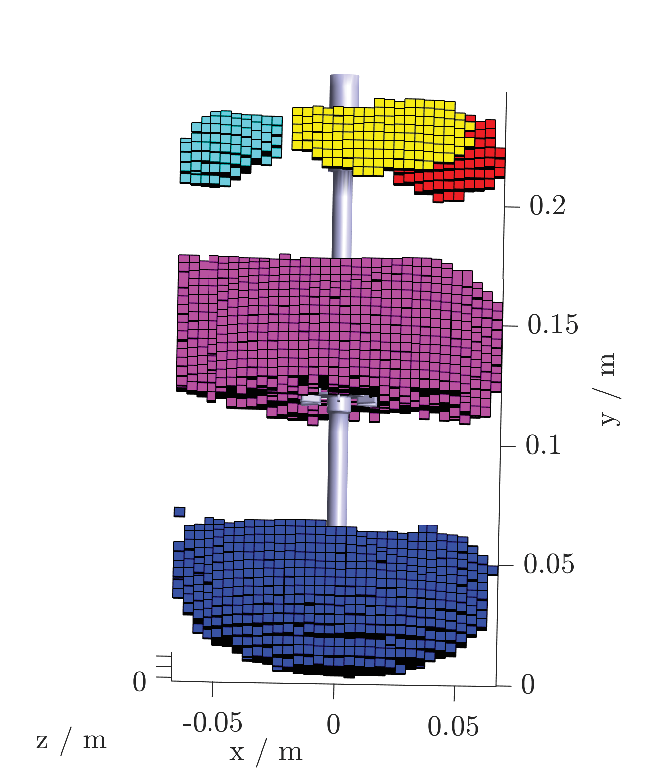} &  \includegraphics[width=0.55\textwidth]{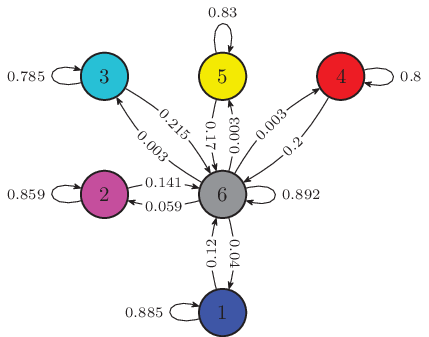} \\
  {\scriptsize(c)} & {\scriptsize (d)} \\
\end{tabular}
\caption{Five extracted almost-invariant sets for the simulated data (a) and experimental data (c) computed from the combined transition information for one stirrer rotation. Markov state compartment model using the five almost-invariant sets and the surrounding region as compartments for the simulated data (b) and experimental data (d). The transition diagrams show the transition probabilities between the compartments over one rotation of the stirrer.}\label{fig:transitionstatediagram}
\end{figure*}

Assuming the dynamics to be nearly periodic with a period equal to one stirrer rotation, we treat the system as time-periodic, following the approach outlined in section \ref{sec:numapproxTO}. In other words, we now combine the information of ten stirrer rotations obtained from the simulated trajectories into a single transition matrix for one stirrer rotation, where 49,983 boxes are used. The actual number of trajectories  in each box at initial times ranges from 1 to 133 with median 77.
Again, we compute the leading eigenvalues (Figure \ref{fig:simexpdata}) and corresponding eigenvectors. The sign structure of the second eigenvectors always divides the reactor into an upper and lower part (shown in Figure \ref{fig:eigenvec}(a)). Subsequent eigenvectors contain information about smaller almost almost-invariant sets, see Figure \ref{fig:eigenvec}(b).
The eigenvectors of the five leading eigenvalues are again post-processed with SEBA (entry-wise maximum of all output vectors in Figure \ref{fig:eigenvec}, as described in section \ref{sec:extractAIS}).
We obtain the $K=5$ almost-invariant sets for the numerical data shown in Figure \ref{fig:transitionstatediagram}(a). The bottom compartment consists of 8,271 boxes, the center of 8,859 boxes, and the three in the top of 431 (cyan), 793 (yellow) and 637 (red) boxes. 

To obtain a simplified representation of the macroscopic dynamics in the tank reactor, we use these five sets (1-5) -- along with the surrounding region (6) -- as compartments in a Markov state model. Since we defined these compartments using the transition probabilities between the boxes, we can also directly obtain the corresponding transition-state probabilities between the six compartments using equation \eqref{eq:coarseP}. The transition-state diagram for the numerical data is shown in Figure \ref{fig:transitionstatediagram}(b).
The transition probabilities in the diagram are low between the different compartments, which means that there is little transport between them. At the same time, the probabilities for the fluid or tracer to stay within its current compartments is very high. This is another demonstration that our identified compartments are almost-invariant.

For the experimental data, not all tracked particles are available at every time frame. To apply our method, we require particles that are available at both the initial and final times of the selected time interval.
Here we can use between 29,856 and 32,831 trajectories for each rotation. 
We again combine consecutive stirrer rotations into a single transition matrix (here we can use all 24 rotations). The reactor domain is partitioned into 44,840 boxes with side lengths $(l_{Bx},l_{By},l_{Bz})=(4.2,3.6,4.2)$\,mm.
The actual number of  trajectories in each box at initial times ranges between 1 and 42 with median 18. We add diffusion using an $\epsilon$-ball with radius $\epsilon = \qty{8}{\mm}$, discretized by 360 grid points. When choosing a smaller diffusion radius (e.g.\ $\epsilon = \qty{4}{\mm}$ as above for the simulated data), the transition matrix has multiple eigenvalues 1, indicating several isolated boxes. Therefore, we use a higher diffusion radius to obtain an eigenvalue $\lambda_2<1$.

When comparing the experimental and simulated data, we observe similar spectral gaps after the second and fifth eigenvalues (Figure \ref{fig:simexpdata}) as well as another, less pronounced eigengap after $\lambda_9$. For the experimental data, we extract $K=5$ very similar almost-invariant sets that are only slightly shifted, with minor differences in sizes (Figure \ref{fig:transitionstatediagram}(c)). Here, the bottom compartment consists of 8,024 boxes, the center of 10,133 boxes, and the three in the top contain 470 (cyan), 401 (yellow) and 540 (red) boxes, respectively. This shows a consistent compartmentalization of the dynamics across numerical and experimental data. The corresponding transition-state diagrams also agree very well (Figure \ref{fig:transitionstatediagram}(d)).

\begin{figure*}[htb]
\centering
\begin{tabular}{@{}cc@{}}  \includegraphics[width=0.5\textwidth]{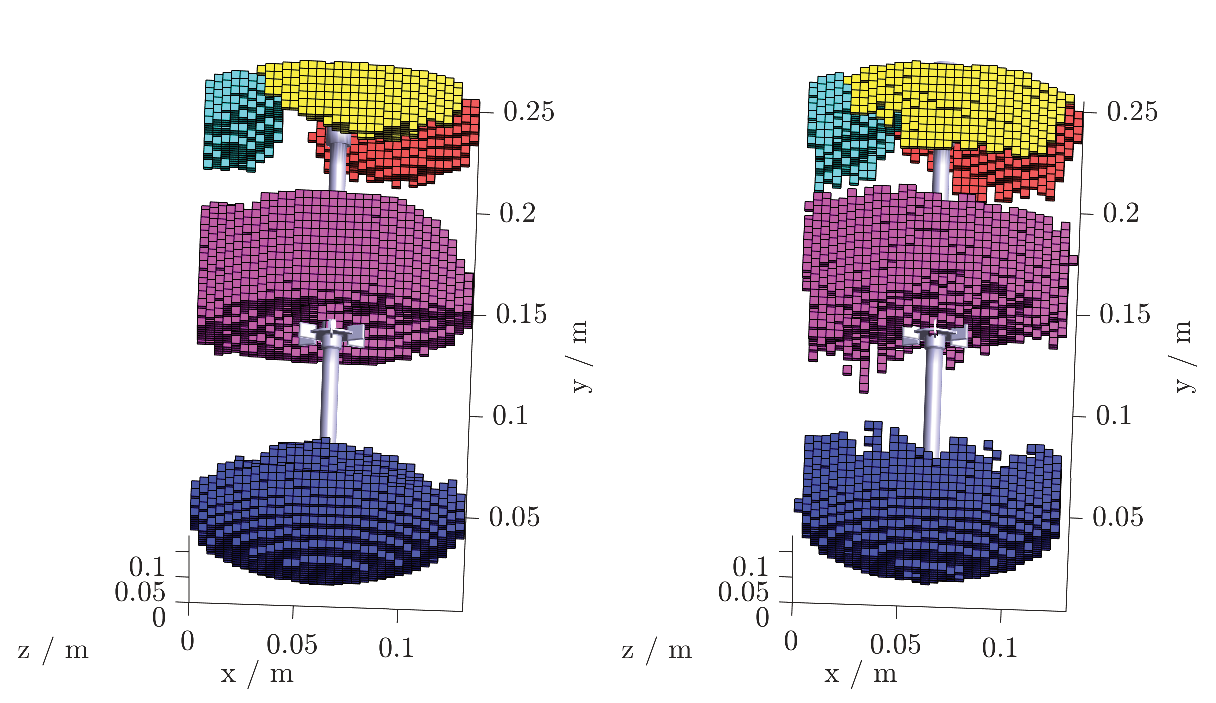} &  \includegraphics[width=0.45\textwidth]{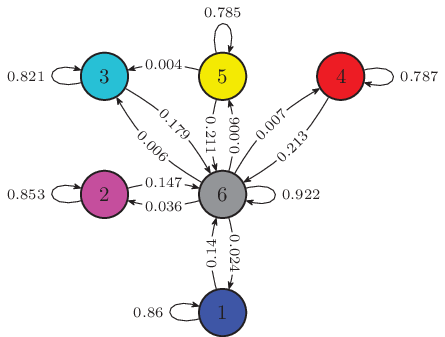} \\
  {\scriptsize(a)} & {\scriptsize (b)} \\
\end{tabular}
\caption{(a): Extracted coherent sets from the combined transition information for one stirrer rotation. (b): Corresponding transition-state diagram for one rotation with the coherent pairs and the background forming the compartments.} \label{fig:coherentsets}
\end{figure*}

In theory, we can also identify compartments that move and change over time. Figure \ref{fig:coherentsets}(a) shows extracted coherent sets for the simulated data using the previously considered single transition matrix for one stirrer rotation constructed from the simulated trajectories given for ten stirrer rotations. For the computation of the coherent sets we used Algorithm \ref{appendix:CS} in the appendix, which utilizes singular vectors of the transition matrix. A transition-state model is derived using equation \eqref{eq:coarseP_time}, see Figure\ref{fig:coherentsets}(b). Note that the compartment pairs in Figure \ref{fig:coherentsets}(a) differ only slightly in shape and position at initial (left) and final times (right). This confirms that our previous assumption of a periodic flow with spatially fixed compartments obtained from almost-invariant sets was reasonable. Very similar transition probabilities in the respective Markov state models (compare Figures \ref{fig:transitionstatediagram}(b) and \ref{fig:coherentsets}(d)) underline this.

From the transition probabilities shown in Figure \ref{fig:transitionstatediagram}(b,d) and Figure \ref{fig:coherentsets}(b) also the physical volume flow rates or material fluxes can be derived. Therefore, the results presented in this section enable us to derive a reactor network for compartment modeling (CM) directly from Lagrangian data. This ensures that the correct dynamical system for the transported quantities forms the basis for the compartments. We stress that the dynamical system of the mean flow, which is typically used for setting up macroscopic models, might be a very different one \cite{Haller2015}. Whether a compartment should then be modeled in a CM by an ideal plug flow reactor or an ideal STR could further be estimated by the internal mixing of the corresponding compartment, which can also be evaluated using our Lagrangian methods. In the future, reactors can be analyzed based on the Lagrangian method presented here, to provide a CM digital twin. This digital twin could then be used for the evaluation of different reaction pathways without the need of further costly experiments or tedious CFD simulations.
\subsection{Mixing statistics}
The computed transition matrix $\bm{P}$ can further be utilized to compute various mixing statistics. Here, we use only the combined single transition matrix for one stirrer rotation of the simulated data to demonstrate these methods.

\begin{figure*}
\centering
\begin{tabular}{@{}cc@{}}
\includegraphics[width=0.4\textwidth]{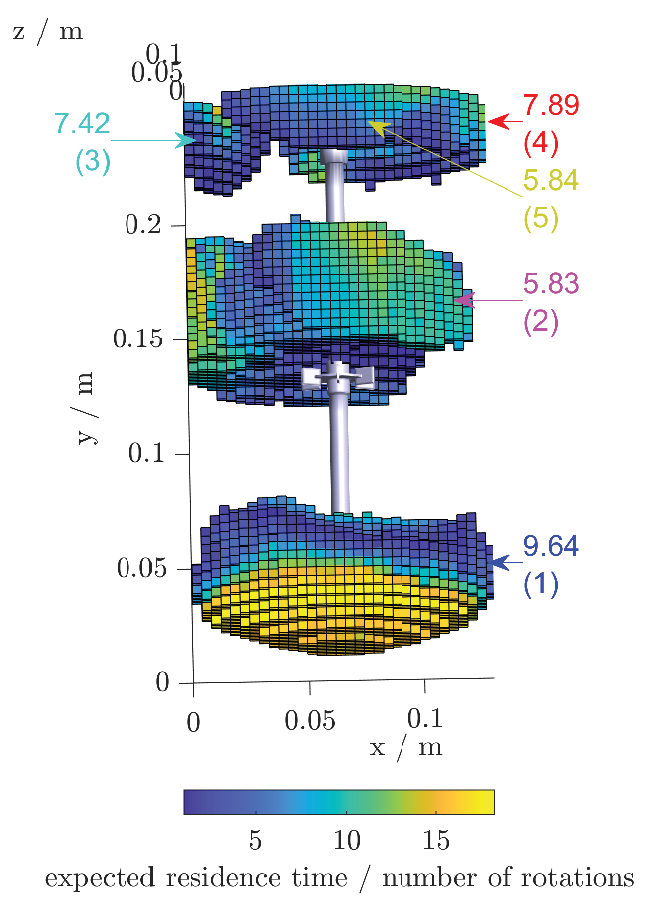} & \includegraphics[width=0.55\textwidth]{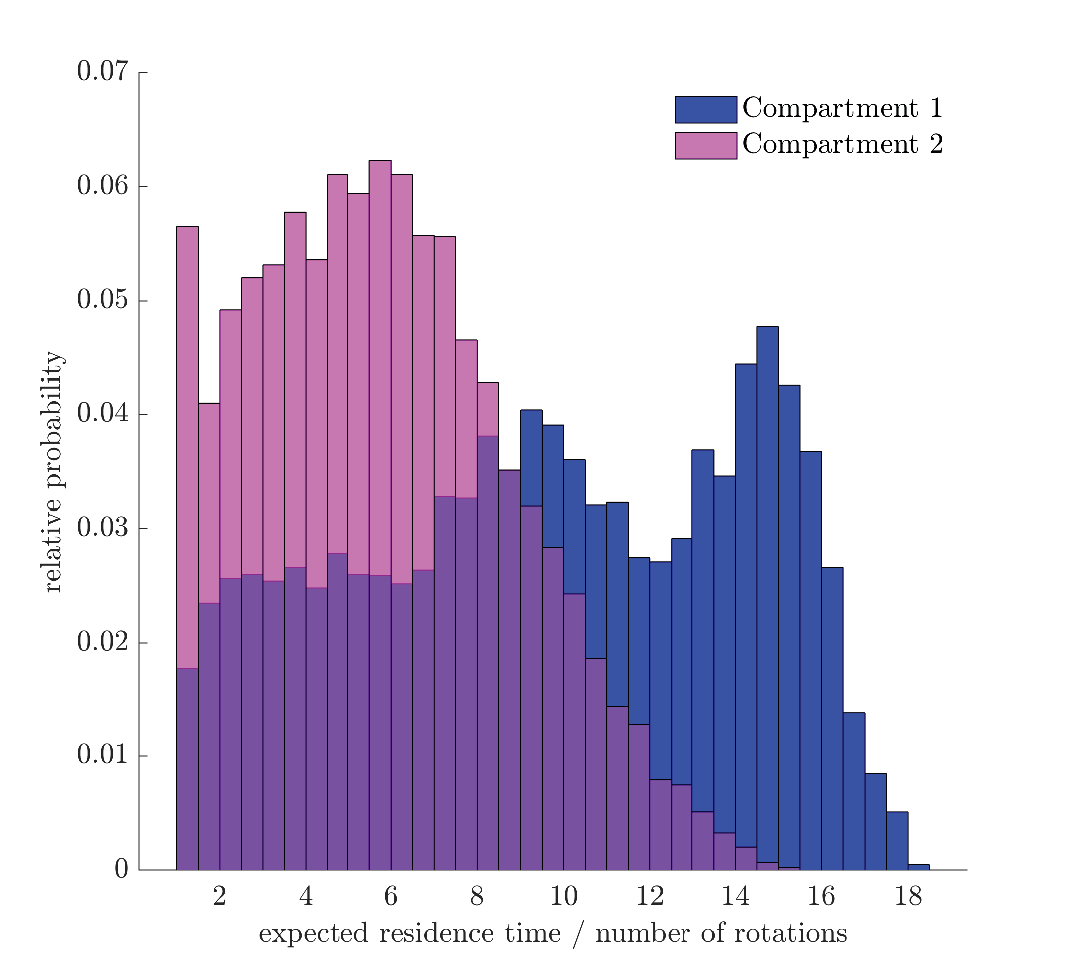} \\
  {\scriptsize (a)} & {\scriptsize (b)} \\
\end{tabular}
\caption{Expected residence times computed from the combined transition information for one stirrer rotation. (a): Color of a box shows the expected residence time for a tracer starting in that box to stay in its initial compartment. The mean expected residence times for compartments (1)-(5) are labeled (in units of stirrer rotations). (b): Histograms of the expected residence times for the middle (magenta) and bottom (blue) compartment. } \label{fig:ert}
\end{figure*}

For each almost-invariant compartment we compute the expected residence times using equation \eqref{eq:SLEresidencetimes} (Figure \ref{fig:ert}(a)).  This quantity represents the average number of rotations required for a particle in a given box to leave the corresponding compartment. The mean expected residence time for each compartment is highest for the bottom compartment with 9.64 rotations, followed by two of the top sets with 7.89 and 7.42 rotations, respectively. The remaining top and the middle compartment both have lower expected residence times with around 5.8 rotations. Notable, a larger proportion of boxes in the bottom compartment have higher expected residence times than those in the other compartments (see for instance the comparison with the middle compartment in Figure \ref{fig:ert}(b)).

\begin{figure*}
  \centering
  \begin{tabular}{@{}cc@{}}  \includegraphics[width=0.32\textwidth]{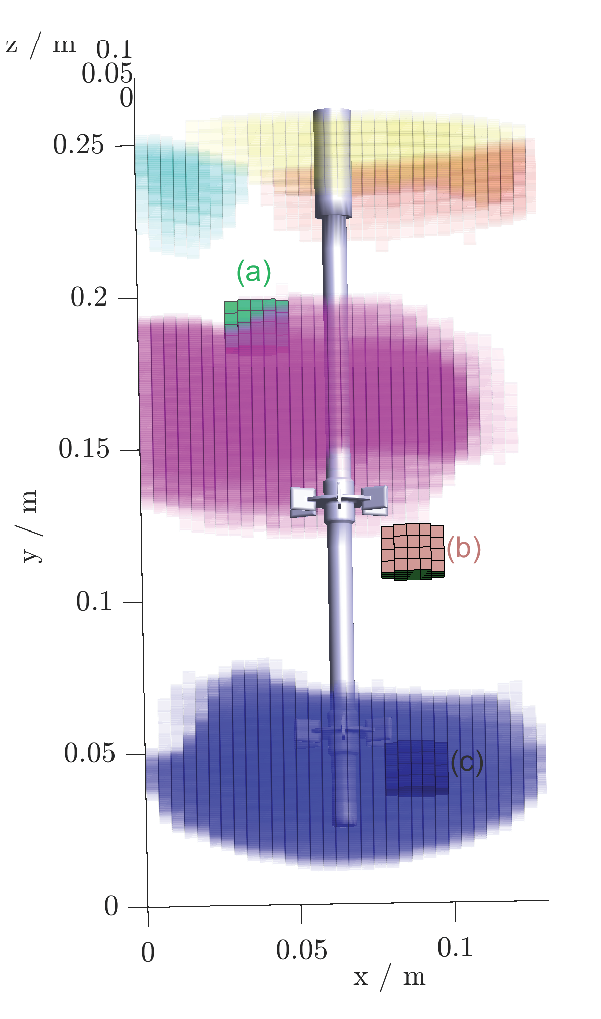} &    \includegraphics[width=0.62\textwidth]{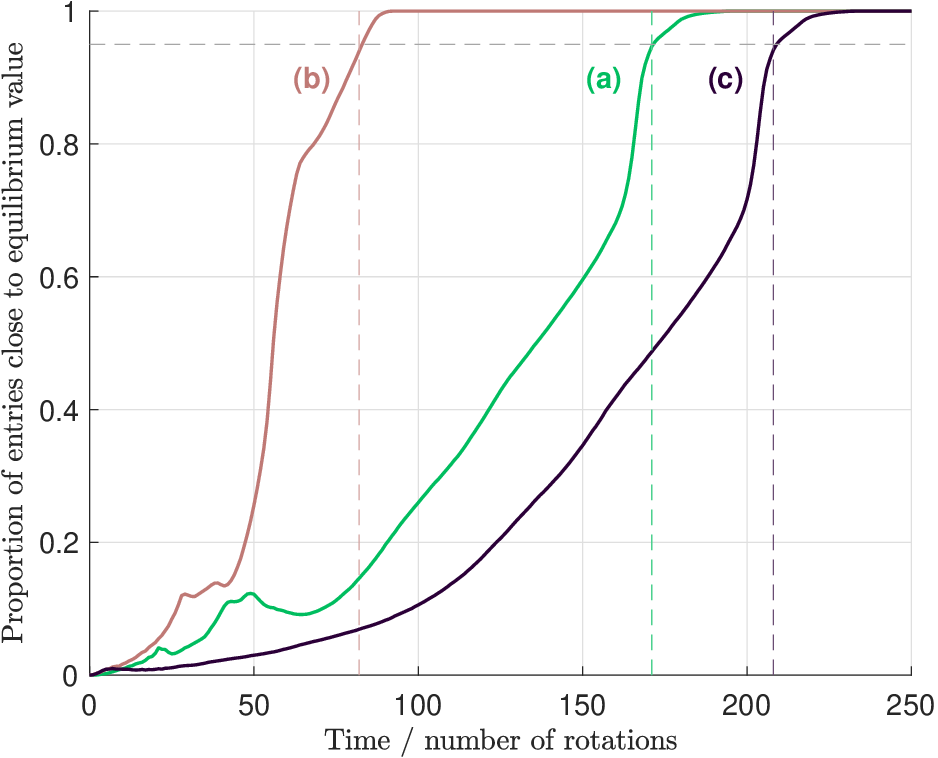} \\
    {\scriptsize (a)} & {\scriptsize (b)} \\
  \end{tabular}
  \caption{(a): Green (spot a), dusty pink (spot b) and dark gray boxes (spot c) show the initialized positive values for the three density vectors. (b): Proportion of vector entries close to the equilibrium value over time for the evolved density vectors. Starting from spot a (green line), spot b (dusty pink line), spot c (dark gray line). The dashed lines mark the mixing times.} \label{fig:mixingtimes}
\end{figure*}

We evolve three different deliberately chosen initial  distributions of a scalar quantity such as a educt concentration. Here, three cylindrical spots are initialized with a uniform density (filled) and zero elsewhere (see Figure \ref{fig:mixingtimes}(a). One is positioned slightly above the middle compartment (spot a), one between the middle and bottom compartments (spot b), and one in the bottom compartment (spot c). Let $\bm{p}_{(x),0}\in \mathbb{R}^n $ (where $n=49,983$ is the number of boxes and $x\in \{a,b,c\}$) be such an initial density vector. This is evolved over time (given in number of stirrer rotations) efficiently  by
$$
\bm{p}_{(x),k+1} = \bm{p}_{(x),k} \bm{P}, \; \; k=0,1, \ldots.
$$
At each time step we compute the proportion of vector entries that are within $\pm 5\%$ of the current equilibrium value, and the mixing time (when 95\% of the boxes are covered) is noted, as described in section \ref{sec:mixingstatistics}. The results are shown in Figure \ref{fig:mixingtimes}(b). 

The region between the bottom and center compartments (spot b) has the fastest mixing time (82 rotations). The other two spots a \& c exhibit much higher mixing times: the top spot (a) has a mixing time more than twice as long (171 rotations), while the spot in the bottom compartment (spot c) has the slowest mixing time (208 rotations). This computational mixing time analysis adds a drastically improved method to perform detailed and simultaneously fast mixing time analysis studies in chemical reactors. Once the transition probabilities are calculated on Lagrangian data, any initial concentration distribution can be analyzed with minor computational effort. Here, the evolution of a density (e.g.\ a vector of concentrations) via sparse matrix-vector multiplications and comparison with the stationary distribution for 250 rotations or time steps (as in Figure \ref{fig:mixingtimes}(b)) is computed in less than \qty{2}{s} on a laptop.
\section{Conclusion}\label{sec:conclusion}
The present study is the first application of classical transfer operator-based computational methods to analyze transport and mixing in a stirred tank reactor. The transfer operator is approximated by a stochastic matrix, where we have estimated the transition probabilities directly from given simulated or experimentally measured Lagrangian trajectory data. Former work has derived compartments for CM models exclusively via non transient, mostly averaged or single snapshot velocity fields \cite{Jourdan2019, decarfort2026,Maldonado2025}, even though, this frozen flow field hypothesis limits the ability of the CM models to capture dynamic
changes in the flow. The method we propose closes this gap in model design and is developed using the full transient Lagrangian dynamics of the fluid.

The transfer operator method itself can be interpreted as a compartment model with the small boxes from the discretization of the domain forming fully mixed compartments. The final Markov-state modeling using Lagrangian coherent flow structures as compartments is an even more coarse-grained CM and reveals the macroscopic transport dynamics. We have motivated the use of almost-invariant or finite-time coherent sets as such coarse-grained compartments, as they could be highly mixing inside but are characterized by little material transport between them.
We find very similar, spatially nearly fixed Lagrangian compartments in experimental and simulated STR data.

Also, the method proposed naturally determines the optimal number and shape of compartments to capture the transient Lagrangian dynamics via the eigengap analysis. This is an advantage in contrast to other methods that try to determine or automatize the optimal set and number of compartments in CFD derived CM models \cite{decarfort2026,Maldonado2025}. For this work, the focus was set on the development of the Lagrangian transfer operator-based method. In the future, however, the methods proposed here can also be directly compared to other methods of compartment modeling since from our considered experimental and numerical data also average velocity fields can be calculated, which thus form a basis for Eulerian based compartment model derivation as, e.g., in \cite{Jourdan2019, laborda2025}. The analysis of the internal mixing inside each compartment could further be used to decide with which reactor type the compartment may be modeled in CM, perfectly mixed or completely segregated.

Further, the transition matrix has been used to very efficiently compute expected residence times and mixing times for different initial concentration distributions mimicking different substrate feed locations. The proposed method is advantageous over methods that need to do elaborate computations of the concentration evolution for each feed location \cite{decarfort2026,Maldonado2025}. Equally, the transition matrix based mixing time and residence time calculation can be used to compare different flow regions and different stirring protocols. We note that the transfer operator approximation in terms of a stochastic matrix induces numerical diffusion of the size of boxes in the discretization of the domain. Moreover, we have included additional diffusion to regularize numerical artifacts resulting from the nonuniform distribution of tracer trajectories. This might speed up the mixing process. Therefore our computed mixing times should be considered as lower bounds.

One limitation of the classical transfer operator method is that a large number of trajectories is required to estimate the transition probabilities to sufficient accuracy. By inducing additional diffusion and cropping trajectories, when there is a certain time-periodicity, we have enriched the trajectory data in the current system and were able to capture the dynamics of the mass transfer.

Modern reactors have more complex geometries and dynamics compared to the classical stirred tank reactor, and the simple assumption of periodic dynamics is no longer possible. In this case, the compartments can be computed by applying the finite-time coherent sets framework. 
However, for sparse data network-based methods for the detection of coherent flow structures from trajectories as already used in a previous STR simulation setting \cite{weiland_computational_2023} might be more convenient.
Moreover, adaptions of this approach can be used to track coherent sets over long time spans \cite{schneide_evolutionary_2022} and to evolve scalar fields in a purely data-based manner \cite{kluenker2025}.
Current work includes extending these methods to the case of experimentally measured trajectories (see \cite{steuwe2026} for a first study).

To optimize chemical reactors a better understanding of the fluid dynamical processes determining transport and mixing is crucial as the local reaction rates depend heavily on the local concentrations of the reactants and our present study is meant to provide a first step in that direction. It delivers a valuable tool to efficiently analyze the aforementioned phenomena and, when coupled with chemical kinetics, to approximate the outcome of a chemical reaction. With the developed methods, different substrate feed positions and reaction pathways can be tested extremely fast using a very simplified but meaningful CM model that bases on Lagrangian transport and therefore captures the true system dynamics. To incorporate chemical reactions we extend the method proposed here in \cite{goebel2026}, where we fuse the evolution of density vectors by numerical transfer operators with update schemes involving chemical reaction kinetics. We also plan to extend these ideas to the trajectory-based setting. 

Although the methods presented in this work are exclusively applied to an STR, they are not restricted to this kind of apparatus and can be expanded to different types such as tubular reactors equipped with lattice structures and flows close to active boundaries. Furthermore, also scale-down models of industrial reactors that are of interest in bioreactors \cite{Gaugler2023} can readily be studied with regard to concentration heterogeneities and mixing times with the proposed methods.

Finally, we aim to develop neural network approaches in order to analyze and quantify transport and mixing from very sparse trajectory data, for instance provided by Lagrangian sensors in industry-scale reactors. The ability to assess the internal state of the reactor in real time (e.g.\ with respect to mixing, dead zones etc.) would be a crucial piece in the development of SMART reactors in the course of the Collaborative Research Centre 1615.

\section*{Acknowledgements}
This project is funded by the Deutsche Forschungsgemeinschaft (DFG, German Research Foundation) – SFB 1615 – 503850735. The experimental data were achieved with the major instrumentation MUST at the HAW Hamburg funded by DFG – 514139948.

\section*{Data availability} 
The simulated STR data are available at \url{https://doi.org/10.15480/882.16844} \cite{weiland2026data}, the experimental STR data will be made available on an open data platform upon publication of this manuscript.

The MATLAB code used for the numerical studies is available at \url{https://gitlab.gwdg.de/anna.kluenker01/compartment-paper_transfer-operator} for peer review purposes.
(The repository will be archived on Zenodo upon publication of this manuscript.) 
The code requires the MATLAB-toolbox for the GAIO library \cite{GAIO} (available at \url{https://github.com/gaioguy/GAIO}).

\bibliographystyle{unsrt}
\bibliography{biblio}

\appendix
\section{Set-oriented algorithms for almost-invariant and finite-time coherent sets}\label{sec:appendix_algorithms}
The following algorithm \ref{appendix:AIS} for finding almost-invariant sets has been adapted from \cite{froyland_padberg_2014almost}, where algorithm \ref{appendix:CS} for approximating finite-time coherent pairs on $M$ has been adapted from \cite{froyland_santi_monahan_10}.
\subsection{Set-oriented approximation of almost-invariant sets}\label{appendix:AIS}
\begin{enumerate}\setlength\itemsep{0.2em}
  \item Partition the state space $M$ into a collection of connected sets $\{B_1,\ldots,B_n\}$ of equal size.
  \item Construct the Ulam matrix $\bm{P}$ using \eqref{eq:pij_approx} (or the data-based versions \eqref{eq:pij_approx_traj},\eqref{eq:pij_approx_traj_diff}), and compute the (assumed unique) fixed left eigenvector $\bm{p}$ of $\bm{P}$.
  \item Construct the matrix $\bm{\hat{P}}$ where $\hat{\bm{P}}_{ij}=\frac{p_j P_{ji}}{p_i}$ and $\bm{R}=(\bm{P}+\bm{\hat{P}})/2$. 
  \item Compute the $K$ largest nontrivial eigenvalues $1>\lambda_2\geq \ldots \geq \lambda_K$ of $\bm{R}$ and corresponding
    eigenvectors $\bm{v}_2, \ldots, \bm{v}_K$.
  \item Extract almost-invariant sets from $\bm{v}_2, \ldots, \bm{v}_K$.\\
\end{enumerate}
\subsection{Set-oriented approximation of finite-time coherent sets}\label{appendix:CS}
\begin{enumerate}\setlength\itemsep{0.2em}
  \item Partition the domain $M$ into a collection of connected sets $\{B_1,\ldots,B_n\}$ of equal size.
  \item Set $\bm{p}=\frac{1}{n}\bm{1}$, where $\bm{1}$ is the all-ones vector as the volume measure to be tracked.
  \item Construct the Ulam matrix $\bm{P}$ using \eqref{eq:pij_approx} (or the data-based versions \eqref{eq:pij_approx_traj},\eqref{eq:pij_approx_traj_diff}), and compute $\bm{q}=\bm{p}\bm{\bar{P}}$. 
  \item Define diagonal matrices $(\bm{D}_{\bm{p}})_{ii}=p_i$ and $(\bm{D}_{\bm{q}})_{ii}=q_i$, $i=1,\ldots,n$,
    compute the $K$ largest nontrivial singular values $1>\sigma_2 \geq \ldots \geq \sigma_K$ of $\bm{D}_p^{1/2} P \bm{D}_q^{-1/2}$ and corresponding left and right singular vectors $\tilde{\bm{v}}_j, \tilde{\bm{w}}_j$, $j=1, \ldots, K$.  Set $\bm{v}_j:=\tilde{\bm{v}}_j\bm{D}^{-1/2}_{\bm{p}}$, $\bm{w}_j:=\tilde{\bm{w}}_j\bm{D}^{-1/2}_{\bm{q}}$.
  \item Extract finite-time coherent sets from $\{\bm{v}_2, \ldots, \bm{v}_K\} $ and $\{\bm{w}_2, \ldots, \bm{w}_K\} $.
\end{enumerate}

\noindent Steps 5.) in both algorithms address the extraction of sets from the leading eigenvectors. We use SEBA \cite{froyland_sparse_2019} for this as outlined in section \ref{sec:set_oriented} in order to obtain $K$ almost-invariant or finite-time coherent sets plus the remainder of the domain as the $(K+1)$-th set. 

\newpage

\begin{center}
\includegraphics[height=1.75in]{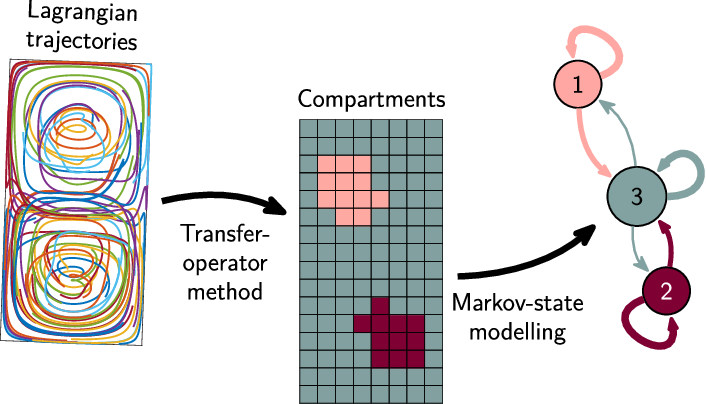}\\[2mm]
Graphical abstract.
\end{center}

\end{document}